# Evidence for length-dependent wire expansion, filament dedensification and consequent degradation of critical current density in Ag-alloy sheathed Bi-2212 wires

A Malagoli[1*], P J Lee[1], A K Ghosh[2], C Scheuerlein[3], M Di Michiel[4], J Jiang[1], U P Trociewitz[1], E E Hellstrom[1] and D C Larbalestier[1]

[1]Applied Superconductivity Center, National High Magnetic Field Laboratory, 2031 E Paul Dirac Dr, Tallahassee, FL 32310, USA
[2]Brookhaven National Laboratory, Upton, NY 11973, USA
[3]European Organization for Nuclear Research (CERN), CH-1211 Geneva, Switzerland
[4]European Synchrotron Radiation Facility (ESRF), 6 rue Jules Horowitz, F-38043 Grenoble, France

E-mail: andrea.malagoli@spin.cnr.it

**Abstract**

It is well known that longer Bi-2212 conductors have significantly lower critical current density ($J_c$) than shorter ones, and recently it has become clear that a major cause of this reduction is internal gas pressure generated during heat treatment, which expands the wire diameter and dedensifies the Bi-2212 filaments. Here we report on the length-dependent expansion of 5 to 240 cm lengths of state-of-the-art, commercial Ag alloy-sheathed Bi-2212 wire after full and some partial heat treatments. Detailed image analysis along the wire length shows that the wire diameter increases with distance from the ends, longer samples often showing evident damage and leaks provoked by the internal gas pressure. Comparison of heat treatments carried out just below the melting point and with the usual melt process makes it clear that melting is crucial to developing high internal pressure. The decay of $J_c$ away from the ends is directly correlated to the local wire diameter increase, which decreases the local Bi-2212 filament mass density and lowers $J_c$, often by well over 50%. It is clear that control of the internal gas pressure is crucial to attaining the full $J_c$ of these very promising round wires and that the very variable properties of Bi-2212 wires are due to the fact that this internal gas pressure has so far not been well controlled.

PACS: 74.25.F-, 74.25.Sv

---

[*] Permanent address: CNR-SPIN, Corso Perrone 24, I-16152 Genova, Italy

## 1. Introduction

Interest in the magnet applications of Bi-2212 ($Bi_2Sr_2CaCu_2O_x$) round wire superconductor has significantly increased [1-16] since recent successes in fabricating high magnetic field coils from high-temperature superconductors (HTS). These HTS coils have generated fields significantly higher [1, 17, 18] than the 24 T in solenoids and 16-18 T in dipoles possible with any Nb-based conductor [19-22]. There is much interest in $REBa_2Cu_3O_7$ (RE = rare earth, REBCO) HTS coils because REBCO coated conductor is available and can be wound as-delivered, especially if co-wound stainless steel is acceptable as insulation [23]. However, the coated conductor tape geometry, with its single-filament architecture, has well known drawbacks compared to round, multifilament wires from which magnets are normally made. In contrast, Bi-2212 is a round, multifilamentary wire with isotropic superconducting properties that was demonstrated to have high whole-conductor or engineering critical current density $J_E$ = 260 A/mm$^2$ at 4.2 K and 45 T [24] and can be made into Rutherford cables [14, 25, 26]. As a round multifilamentary wire, Bi-2212 has great potential for high field applications in NMR magnets and accelerators [1, 26-28]. It may also be useful in some special cases, like the multi-circuit HTS current leads needed to connect above-ground power supplies to the magnets in the accelerator tunnels of the LHC at CERN [29].

A serious problem that has limited the application of Bi-2212 so far is that short wire lengths have high $J_c$, while coil lengths have much lower $J_c$, often less than 50% of short sample $J_c$ [5, 30-31]. Recent work has shown that a key constraint on high $J_c$ development in Bi-2212 is the development of gas bubbles that form by agglomeration of the distributed porosity of the Bi-2212 powder during heat treatment. Because simultaneous deformation of the silver matrix and the filament space is only possible by powder sliding, the terminal density of the powder is typically only 60-65% [32]. This uniformly distributed porosity becomes highly non-uniform during heat treatment by agglomeration into bubbles which greatly interrupt the Bi-2212 filament connectivity [2, 4, 6-7]. Gas trapped in the filament space, which can include residual air as well as $CO_2$ and $H_2O$ from impurities in the wire, expands, causing the wire to increase in diameter and the Bi-2212 filaments to dedensify. In wires with open ends, some gas can escape from the wire ends while it is being heated, reducing the trapped bubble density, the filament dedensification and consequent loss of $J_E$, without, however, fully eliminating the issue. Bubbles can be almost removed from short wires (~8 cm) by pre-heating open-ended wires in pure $O_2$, cooling to room temperature, cold isostatically pressing (CIP) the wire to remove most of the 30-35% residual porosity in the filament space, and then applying the standard heat treatment to the oxygen-containing open-ended wire [7]. This densification process doubles the overall conductor current density $J_E$ to 800 A/mm$^2$ (the superconducting filament current density $J_c$ = 3600 A/mm$^2$) at 5 T, 4.2 K [7], which is equivalent to $J_E$ = 550 A/mm$^2$ ($J_c$ = 2400 A/mm$^2$) at 20 T. Such $J_E$ values are very attractive for both solenoid and dipole magnet designs, making understanding of the dedensification problem essential to ensuring that round wire Bi-2212 becomes a viable conductor for magnet technology.

An earlier experiment of ours showed a substantial dependence of $I_c$ with distance from the ends of 1 m long wires [5], and a direct correlation between the local $I_c$ and small variations of the local mass density of the wire. The experiments described here follow on and generate new insights into the cause of the loss of $J_E$ in longer wires. Our new data are also relevant to our earlier measurements on short samples extracted from deconstructed coils, which indeed showed low $J_E$, but also very little $J_c$ and $J_E$ length dependence [10]. On the other hand, a strong length dependence was observed in the frequency of Bi-2212 leaks, which clearly increased towards the middle of the coil and away from the ends of the 278 m long conductor. The lack of $J_c$ and $J_E$ length dependence in coils for samples cut several meters from each end suggests that the onset of "long wire" behavior occurs in the 5-10 m range and emphasizes that the effects reported in this paper are only expected to be seen at shorter lengths or the very ends of coils. The key result which unifies the two data sets is the fact that the extracted coil $J_E$ samples are all strongly degraded, no doubt due to a general wire diameter expansion which, however, was not measured since the experiment ended before this one started. The value of correlating low $J_E$ or $J_c$ to the easily measured wire diameter is indeed a vital recommendation of this paper because it is now clear that the fundamental cause of degraded $J_E$ is due to filament dedensification driven by the internal gas pressures during heat treatment.

The purpose of the present study was to explore when the key pressure generation step occurred, to explore the length dependence of the critical current properties so as to understand what might be the best choice of "short sample", and to contrast the behavior of open-ended wires with closed-end wires that would prevent gas from escaping the ends of the wire and thus mimic the behavior of the long wires found in coils. Our aim was to make new measurements of the structural and superconducting properties as a function of the sample length and to explore new techniques that would allow rapid and effective measurement of wire expansion, while also giving more data than conventional metallography of wire cross sections. We used three techniques in this study, in addition to conventional metallography. We used (1) a flat-bed scanner of high optical resolution for routine, detailed length-dependent diameter measurements on any wire that could be laid flat, (2) a laser micrometer to give a more accurate and higher resolution diameter measurement check and (3) diffraction and tomography synchrotron X-ray analysis to provide detailed internal phase and bubble structure of a few specially selected regions. All showed a strong correlation between degraded local $J_c$ and local diameter expansion. Collectively our results make it clear that it is vital to better understand and control the dedensification caused by internal wire pressure if Bi-2212 wire technology is to reach its full potential.

An additional outcome of our study is that it provides an answer to the question of what constitutes a representative short sample for evaluation of new wires. Our experiments show that the $J_c$ of 5-10 cm long, open-ended wires is much less degraded by the internal pressure that dedensifies the Bi-2212 than longer wires. Thus short, open-ended wires are more useful for evaluating the full potential than longer wires. However, such samples are still not fully dense and their Bi-2212 filament array is thus still not optimally connected, as shown by earlier studies on CIPped wires where $J_E$ was doubled by densification [7]. Our studies suggest that any length of presently-available, as-drawn wire heat treated with well-sealed ends will expand during heat treatment in a way representative of long wires wound into coils, dropping $J_E$ by factors of 3 or more. We conclude that there is no clear way to understand the full critical current performance of Bi-2212 wires unless the wire is fully densified, a topic that has been addressed by work still to be reported [35].

## 2. Experimental details

Several lengths of the same Bi-2212 round wire made at Oxford Superconducting Technology by the powder-in-tube (PIT) method [33] were analyzed. A pure silver tube was filled with Bi-2212 powder made by Nexans and then drawn into a hexagonal form. In the present conductor, 37 such hexagonal filaments were restacked inside a pure Ag tube and again drawn into hexagons. 18 of these first-stage 37-filament bundles were restacked in a Ag-0.2 wt % Mg alloy tube and then drawn to 0.8 mm diameter. We analyzed samples that were 5, 15, 100, and 240 cm long after a standard melt process heat treatment in 1 bar flowing $O_2$ [7] whose maximum temperature was 888°C so as to melt the 2212 powder. We ensured that the whole length of each sample was within the 30 cm long homogeneous zone of the furnace (±0.5°C) by bending the 100, 150, and 240 cm long samples into the hairpin shape shown in Figure 1. The ends of the open-ended wires were polished to expose each of the filaments to the ambient furnace atmosphere. We closed the ends of some wires by dipping them into molten Ag.

We used four different techniques to evaluate the length-dependent wire diameter after heat treatment. Longitudinal, and transverse, polished cross-sections for conventional metallography were made by using a scanning laser confocal microscope (SLCM, Olympus LEXT OLS3100) that gave high contrast digital images with a resolution of ~0.12 μm [5]. The transverse cross section images were analyzed with ImageJ [34] to extract the total area and the fraction of Ag matrix and filament pack to evaluate the sample diameter and the area occupied by the filaments. These techniques provided length-dependent microstructures and diameters with a length frequency of about one every few cm. To obtain more frequent diameter data on smaller length scales, we employed a 350 mm long scanning bed (Agfa Duoscan 2500) with an optical resolution of 2500 dpi (1 pixel = 10.2 µm) which allowed us to digitize the full length of each sample while it was still in one piece with a projected diameter resolution of about 10 µm. An advantage of this scanner is that it has a good depth of field, which is important because the maximum diameter of this wire is always at least 0.4 mm above the scanner bed and it can be even higher if the wire does not lie flat. Two other flatbed scanners with higher advertised resolution were tested (Epson Perfection Photo series V600 and V700) but they were found to have lower effective resolution at half-strand diameter depth. Full reproducible scanner control was obtained by using VueScan software. Image processing and thresholding were performed using Fovea Pro and ImageJ software, and an Excel macro was used to calculate projected diameters along the wire length. The flatbed scanner also provides images of the wire surface that can be used to identify defects.

After imaging with the flatbed scanner, some of the hairpin samples were cut into several pieces, the exact location of these pieces being recorded. Their diameters were measured using a static 8-axis laser micrometer [BETA LaserMike, 183B-100E] with a resolution of about 1 µm. Selected straight samples from the hairpins were also used to measure the mass/unit length, for $I_C$ measurements and for combined synchrotron absorption micro-tomography and X-ray powder diffraction measurements. The X-ray measurements were performed at the ID15A beam line of the European Synchrotron Radiation Facility (ESRF) using a monochromatic 70.0 keV X-ray beam with a bandwidth of 0.7 keV. The image pixel size of these x-ray measurements is 1.194 ×1.194 µm² [4].

Transport $I_C$ measurements were performed in liquid helium at 4.2 K in a magnetic field of 5 T using the four-probe method applying an electric field criterion of $10^{-6}$ Vcm$^{-1}$. The longitudinal axis of the wire was perpendicular to the magnetic field direction. The voltage taps were spaced 1 to 1.5 cm apart in the center of samples.

## 3. Results

*3.1 5 and 15 cm open-ended wires*

**Table I** shows data for 5 and 15 cm long samples heat treated with open ends. The data show little change in wire diameter after heat treatment and $I_c$(4.2 K, 5 T) values are very similar. Apparently the internal gas pressure can equilibrate to a similar extent in such short samples.

*3.2 1.0 and 2.4 m open-ended wires*

The results for these wires are also summarized in Table I. These open-ended samples also allowed some gas to escape during the heat treatment but to a significantly lesser extent than in the 5 and 15 cm long samples. The $I_c$ – length behavior along such 1 m long samples has already been thoroughly analyzed and reported in a previous paper [5].

**Table I**: Comparison of short and long sample properties in open-ended samples. Wire diameters were measured with the laser micrometer unless otherwise noted.

| Reacted sample length | Wire diameter [mm] | End to end $I_c$(4.2K, 5T) range [A] |
|---|---|---|
| As-drawn before reaction | 0.802 | 0 |
| 5 cm | 0.804 | 200 - 210 |
| 15 cm | 0.803 – 0.804 | 200 - 210 |
| 1 m | 0.812 - 0.820 | 155 - 210 |
| 2.4 m | 0.81 – 0.85<br>0.83 – 0.85** | 70-180 |

\*\*Wire diameter measured with the flat-bed scanner.

Figure 1 shows images of the 1 and 2.4 m open-ended samples after full heat treatment. The 1 m wire did not leak, whereas the 2.4 m wire leaked at one of the open ends and at a few points along the wire length. Table I shows that $I_c$ and the wire diameter varied significantly in the 1 and 2.4 m wires. Figure 2 shows how $I_c$ and wire diameter varied along the length of the 1 m open-ended wire. The diameter at the ends of the 1 m wire was ~0.812 mm, about ~1% larger than for the 5 and 15 cm wires. The diameter increased to ~0.82 mm in the middle of the wire, ~1% greater than at the ends. $I_c$ at the ends is ~210 A, the same as in the 5 and 15 cm samples but it decreased by ~25% to ~155 A in the middle of the 1 meter long wire.

Figure 3 shows $I_c$, diameter and mass/length as a function of position in the 2.4 m open-ended wire. The $I_c$ at the ends of the wire was ~180 A, ~14% lower than in the 5 and 15 cm and 1 m wires, but it dropped sharply to only ~70 A in the middle of the wire, only ~33% of that seen in the 5 and 15 cm samples, while the change of mass/length is only about 1 %. It is clear therefore that even open-ended wires longer than 15 cm quickly show strong $I_c$ degradation and that this degradation is associated with significant wire diameter expansion.

Figure 4 shows the projected diameters of the straight sections of the as-drawn and heat-treated 2.4 m wires measured using the flatbed scanner. The diameters of the as-drawn wire are relatively uniform, with no more than ~±2% variation. In contrast, after heat treatment, the diameter of the entire wire increased relative to the as-drawn wire, especially towards the middle of the sample. Moreover, the ends of the heat treated 2.4 m wire have larger diameters than the ends of the heat treated 1 m wire (Figure 2b). Figure 4 also shows localized diameter spikes in the heat treated wire, some of which are shown in Figure 5. After the heat treated wire was scanned, it was cut into 8 straight and 7 curved sections. The diameter of the straight sections was measured using the laser micrometer and these data are also included in Figure 4. The laser micrometer data have the same general shape as those from the flatbed scanner; however, the scanner diameters are generally 10-30 μm (~3%) larger than those from the laser micrometer, possibly due to an error in the absolute calibration of the scanner. Figure 5 shows images in regions where there was a diameter spike produced both by the scanner and by imaging with the SLCM, as one of the insets in Figure 5 shows. Once optimum settings were determined, the flatbed scanner was fast, inexpensive, and non-invasive, and required little sample preparation. Although it has a resolution only of ~10 μm, it easily detected the wire swelling towards the middle of each sample, as well as bursts and cracks, and it also provided images of the wire surface at these locations. The high reflectivity of unreacted strands made them more difficult to measure than the heat-treated strands.

Several samples for synchrotron X-ray tomography imaging were chosen from the straight sections of the 2.4 m open-ended wire, some of which are shown in Figures 6 and 7. These images reveal the internal structure within end and mid-point wire sections and in a leaked region. The two transverse cross sectional tomograms in Figure 7 directly show the larger diameter of the midpoint wire section compared to the wire end. In addition, the midpoint tomogram has darker gray filaments because it has more filament void space than at the ends of the wire. The leaked section shows severe internal damage consistent with the generation of large internal pressures. We note that Figures 6 and 7 are through-sample tomographic X-ray sections that avoid all artifacts that might be produced by metallographic sectioning where both filament pull out and void filling are perennial problems to be addressed in the interpretation of polished cross-sections. Since the tomography is completely non-invasive, all the voids and filament damage zones imaged in these tomographs are real, and it is clear that the voids are extensive at all points along the wire.

Figure 8 shows the gray-scale histograms for tomograms collected at different points along the wire. In order to exclude void space outside the wire from the void estimates, it was necessary to crop a part of the outer Ag sheath. The selected region of interest has a radius of 377 μm. The grey scale is in arbitrary units, a higher grey value indicating a stronger attenuation of the 70 keV X-rays. For reference, the histogram acquired for a wire in which porosity had been strongly reduced by 100 bar overpressure processing [35] is also shown. The shift of intensity to higher white scales is quite clear, as is the fact that the void fraction increases towards the midpoint of the wire. The massively leaked region clearly has a very large amount of void.

The filament void volume changes along the length of the wire were roughly estimated from these histograms and the results are summarized in Table II using the following assumptions: (1) there are no voids in the overpressure processed wire, (2) all voxels with grey values > 25 are from Ag, and (3) the Bi-2212 volume is constant along the wire to a precision of better than 2-3% as determined from the simultaneously recorded X-ray diffractograms. The rather shocking conclusion from such measurements is that the 2212 filament mass density is easily driven well below 50% after reaction and that major reductions in density occur as the distance from the end of the 2.4 m long wire increases.

**Table II**: Void volume within the filaments estimated from the grey value distributions of Figure 8 using the assumptions given in the text and the effective 2212 density calculated from the filament void volume. Distances are measured from one end of the 2.4 m long wire.

| Sample ID | A8-1 | C7-3 | D2-3 | E6-1 | E6-4 | F3- | G5-3 | H4-2 | E6-4 (big leak) |
|---|---|---|---|---|---|---|---|---|---|
| Distance from one end of the wire (mm) | 30 | 798 | 1060 | 1294 | 1479 | 1780 | 2000 | 2290 | 1479 |
| Void volume (%) | 53 | 63 | 64 | 63 | 62 | 65 | 57 | 52 | 71 |
| 2212 filament density (g/cm$^3$) | 3.1 | 2.4 | 2.3 | 2.4 | 2.5 | 2.3 | 2.8 | 3.1 | |

*3.2.3 **2.4 m long, closed-end wire***

We sealed the ends of a 2.4 m wire to simulate the behavior of a long coil length, since the $J_E$ degradation effect increased so markedly on going from 1 to 2.4 m long wires. Our expectation of course was that gas would be even less able to escape from the wire ends during heat treatment. We were also interested in determining when during the heat treatment the dominant expansion and/or leakage occurs. To do this we heated the 2.4 m closed-end wire twice, first to 875°C for ~72 h just below the 2212 melting point (~882°C), then, after cooling to room temperature and diameter measurement, into the melt state at 888°C. The time of the 1$^{st}$ heat treatment at 875°C was chosen to duplicate the time in the melt of the standard process so that creep of the Ag matrix would be close to that which should occur only 13°C higher when a full melt of the Bi-2212 occurred [3]. Figure 9 shows that this 875°C heat treatment produced only one small leak from one end and no leaks along the wire length. The end leak showed us that our sealing technique is sometimes imperfect and that some liquid was present at 875°C, consistent with the earlier synchrotron observations of some liquid before the main 2212 melt temperature [4]. Flat-bed scanner measurements (Figure 9) did show a small 20 to 30 μm diameter increase over the whole wire length, but without local bulges, leaks or preferential diameter increase towards the middle of the wire. A small and rather uniform creep of the Ag under the internal gas pressure evidently occurred.

This closed-end wire was then put through a standard heat treatment going now up to 888°C in order to melt the 2212. Figure 10a shows there was then significant leakage along the entire wire length, dramatically greater than in the fully processed, open-ended wire (Figure 1b) or in the wire before melting (Figure 9). These leaks typically correlate to local wire bulging, as shown in the insets in Figure 10b. Perhaps unexpectedly, the scanner data in Figure 10b show that the additional full heat treatment produced less length-dependent swelling and increased the average diameter a little less than in the open-ended wire, but this is probably because the leak frequency was much greater and thus pressure release occurred more frequently in the closed-end than in the open-ended sample.

## 4. Discussion

Our experiment started as a simple extension of our earlier one [5] on samples up to 1 m in length with the goal of finding a characteristic decay length for $I_c$ and $J_E$ in which it was clear that even 1 m long samples did not accurately reflect the much lower $J_E$ values found in coil length wires [10]. What we actually find here is that there is no characteristic $J_E$ decay length, because $J_E$ is controlled by the local Bi-2212 density which is in turn controlled by the extent to which local internal gas pressure can be relieved, either by diffusion towards the ends or by diameter expansion, bulging or, in the worst case, tearing of the Ag matrix and sheath. Broadly speaking we can say that there is a significant length-dependent depression of $I_c$ or $J_E$ that correlates to increase in the wire diameter after reaction heat treatment, that closed end wires expand more and leak much more than open-ended wires, and that open-ended wires allow partial gas release, even at 2.4 m length, while closed end wires do not. The fundamental cause of the lowered $I_c$ and $J_E$ of long-length wires is the dedensification of the Bi-2212 filaments produced by wire expansion during the melt phase of the heat treatment. Since the drawn wires are never more than 60-65% dense, the final density of Bi-2212 filaments is well below 50%, greatly reducing the superconducting connectivity of the filament bundle. The fact that this dedensification occurs particularly in coil-length wires is why the many desirable characteristics of a round wire, multifilament conductor with upper critical field several times higher than $Nb_3Sn$ have seen so little employment in real high field magnets. We believe that the present experiments point unambiguously to what a new processing technology must do – to prevent filament dedensification and loss of connectivity.

A first question is to ask what kind of short sample can be realistically used to assess the intrinsic capabilities of wires being made today. Table I shows that 5 and 15 cm open-ended samples are short enough that gases can escape from the wire ends during heat treatment, avoiding significant gas pressure build up that could expand the wire diameter. Thus, short, heat treated wires have the same diameter as as-drawn wires and for this reason avoid 2212 dedensification and leakage during heat treatment and have $I_c$ and $J_E$ values representative of 2212 filaments of the same density as the drawn wire. Such properties would be valuable for coil technology but it does need to be recognized that the filament connectivity of such short wires is still significantly compromised because the filaments contain gas bubbles corresponding to at least the residual ~35 vol.% powder void fraction [32, 36]. These bubbles form as the highly distributed gas that fills the pore space agglomerates during melt processing. We believe that these bubbles still significantly limit the Bi-2212 connectivity, as is indicated by the doubled $I_c$ and $J_E$ values obtained by special CIPping and swaging that minimize the pore volume prior to heat treatment [7, 36]. This possibility sets a very positive opportunity for this conductor technology.

One meter long ITER barrel samples are generally used as the standard for evaluating $Nb_3Sn$ conductors and often used for Bi-2212 wire evaluation too. However, the experiments here suggest that 1 m samples are not representative of either short or long wires, even though the ends of open-ended (the standard preparation route) wires have diameters only very slightly (~1%) larger than as-drawn wire, indicating only slight Bi-2212 dedensification there. Indeed the $I_c$ of these ends was the same as for the shorter 5 and 15 cm long samples. However, Figure 2 shows that $J_E$ of the center wire region was very dependent on distance from the end, falling to about 75% of its end value in the center. Thus a 1 meter long sample can provide information both on short sample properties without additional dedensification, and can be a guide to the dependence of $J_E$ on distance from the end if the length dependent wire diameter is also measured.

However, the lack of generality of even the 1 m sample is quite obvious in the 2.4 meter long sample. The diameter of the wire ends was between ~1.5 and ~4 % larger than the as-drawn wire, while $I_c$ of the end sections was ~14% lower than in the ends of the 0.05, 0.15 and 1 m wires. The diameter of the middle of the 2.4 m open-ended wire was ~8% larger than the as-drawn wire and ~6% larger than the middle of the 1 m wire. By contrast to the minimum $I_c$ of 155 A seen in the 1 meter wire, the minimum $I_c$ for the 2.4 meter wire was 70 A, only one third of the end $I_c$ values. In this case, the wire showed many bulges, some to more than 0.9 mm diameter. There clearly is a highly non-linear dependence of $I_c$ on wire diameter, whose particulars depend on the local Bi-2212 connectivity and mass density.

Because most testing of Bi-2212 wires has hitherto occurred with open-ended samples, including coil reactions, we wanted to explicitly confine the internal gas by closing the ends, aiming to produce guidance on the impact on long strand length properties from more manageable and cost-effective short samples. For this 0.8 mm diameter wire, dipping into molten Ag worked reasonably, though not perfectly, to seal the ends. Our recent tests indicate that such samples provide a good worst-case test when the ends are well closed, because we saw much more leakage in the 2.4 meter sample (Figure 10) than in a recently deconstructed 278 m wire length coil [10]. Most leakage did occur at the center of the coil, where we expect the maximum pressure build up, but leaks were tens of cm apart, rather than the few mm separations seen in Figure 10. This does suggest that some gas escape does occur in even hundreds of meters of wire, making it uncertain what kind of properties can be predicted for open-ended coils based on short sample measurements, where the length is arbitrary and the diameter expansion uncontrolled and generally unknown.

The final issue to address is the root cause of the problem. Clearly wire expansion occurs by an internal pressure-driven creep of the Ag sheath, as has been explored recently in more detail by Shen *et al.* [37]. Our purpose in exploring the expansion of a closed-end 2.4 m wire was to see at what temperature the majority of this expansion occurred, especially whether just below the melting point of the Bi-2212 or only after melting at 888°C. A small diameter increase to 0.825 mm did indeed occur during the first heating to 875°C but the diameter measurements (Figure 9) showed no bulges. When the wire was heated to 888°C in the second heat treatment, there was an additional, length-independent average increase in wire diameter to 0.84-0.85 mm, but now with significant local bulging and leakage. We thus conclude that the most significant rise in internal gas pressure occurs only after the onset of melting.

The potential sources of this pressure increase on melting are several. One is from oxygen release when 2212 melts. However, silver is permeable to oxygen, as it is to the oxygen of the residual gas trapped in the void space of the 60-65% dense filaments. Ideal gas law calculations of this residual gas expansion suggest that air at 1 bar at room temperature could develop a 3.9 bar pressure at 875-890°C. Evidently this pressure is not sufficient to cause major Ag creep and 2212 dedensification when the 2.4 m long wire was heated to 875°C before melting, leading us to propose a larger role for condensed phases such as C impurity or moisture that produce gaseous $CO_2$ and $H_2O$ on melting, greatly enhancing the total internal pressure and initiating much greater creep only at the onset of melting. Indeed we measured significant C in this wire (~250 ppm by wt.), well above the starting powder specifications (~40 ppm by wt.). In summary, we conclude that melting is essential to the dedensification that drives down $I_c$, $J_c$ and $J_E$ of long length wires, that the means by which the dedensification occurs is by pressure-induced creep of the Ag sheath, and that gasification of impurities, especially $H_2O$ and C, is an essential component of the internal pressure that drives dedensification. In our view, we believe that length-dependent $I_c$ properties will be quite variable until impurities are brought under better control. Recent overpressure processing experiments conducted in our laboratory which we discuss in future papers do show the value of this approach [35].

## 5. Conclusions

This study clearly shows that wire expansion can be prevented in short samples with open ends where the gas can escape from the ends of the wire during the heat treatment. For wires longer than a few cm, say 15-100 cm, all the gas cannot escape from the ends of the wire during the heat treatment and the middle of the wire expands, which dedensifies the 2212 filaments, decreases the 2212 connectivity and reduces $J_c$. Our experiments with closed-end wires show that the very significant decrease in $J_c$ that occurs in coil-length wires is due to internal gas pressure generated largely on melting the 2212 phase. Our experiments suggest that it would be valuable to fabricate wires so that (1) oxygen is the only gas in the as-drawn wire (because it can pass freely through Ag), (2) to use high purity powder with very low concentrations of condensed phase contaminants like C and $H_2O$, and (3) to prevent impurity pickup, especially C and $H_2O$ during wire fabrication because both can gasify on Bi-2212 melting. These conclusions are all consistent with the earlier experiments of Jiang *et al.* [7, 35, 36] which demonstrate that the highest $J_c$ is only developed when the 2212 filaments are fully densified and all residual gas bubbles are eliminated.

## Acknowledgements

This work was carried out within the Very High Field Magnet Collaboration (VHFSMC) which was supported by an ARRA grant of the US Department of Energy Office of High Energy Physics by the National High Magnetic Field Laboratory, which is supported by the National Science Foundation under NSF/DMR-1157490 and by the State of Florida. We are grateful to many discussions from partners within the VHFSMC collaboration, especially from Tengming Shen who has explored the Ag creep aspects of the problem and to Emanuela Barzi who first drew attention to the expansion of long wires in Rutherford cables that also manifests itself here in short samples. We acknowledge the ESRF for beamtime at ID15.

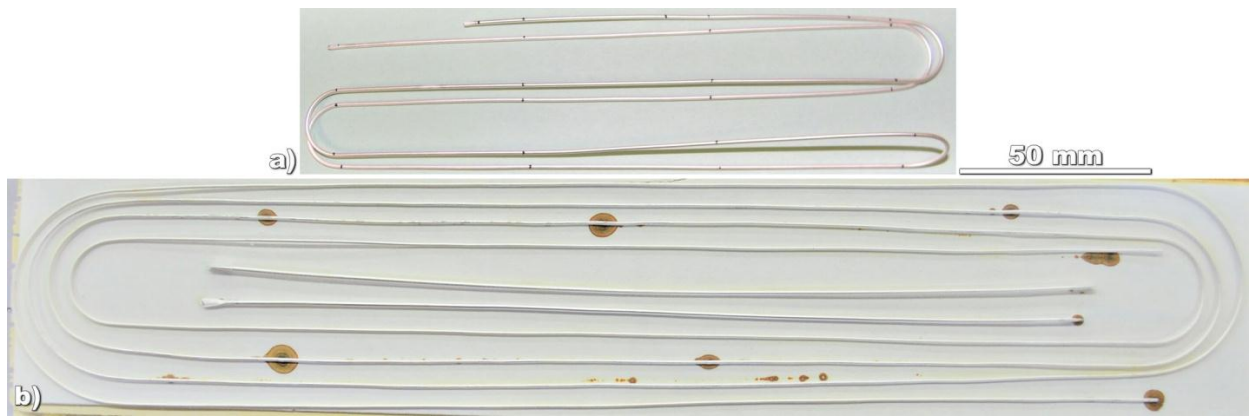

**Figure 1**. Digital camera images of (a) the 1 m open-ended wire and (b) the 2.4 m open-ended wire with two shorter, open-ended straight samples place in the middle of the hair pin. The dark dots on the 1 m wire in (a) are added marks used to define the cutting into short straight sections. In (b) brown stains are seen on the alumina fiber sheet under the strand where leaks generated by internal pressure have occurred during heat treatment. Both large individual leaks and clusters of smaller clusters can be observed on the straight sections of the 2.4 wire.

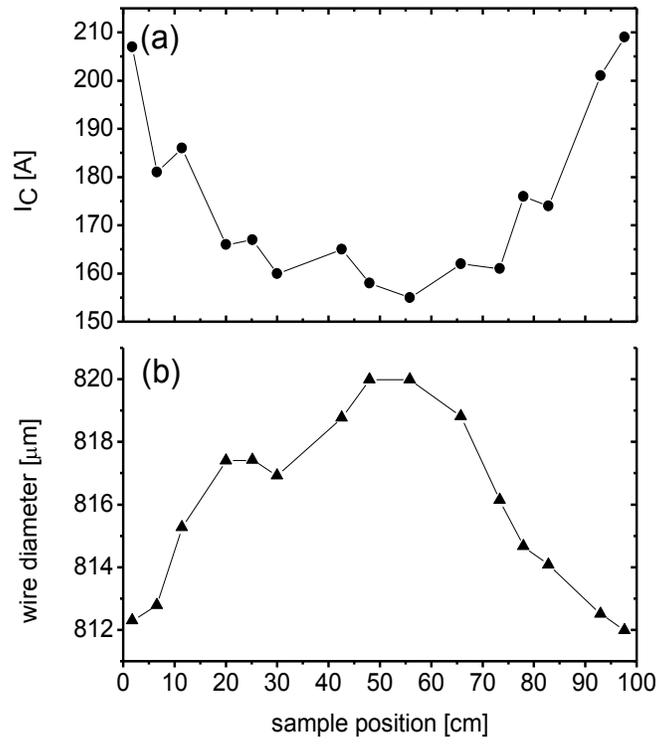

**Figure 2.** The variation of (a) $I_c$ and (b) the wire diameter along the length of the 1 m open-ended wire. $I_c$ was measured at 4.2 K, 5 T.

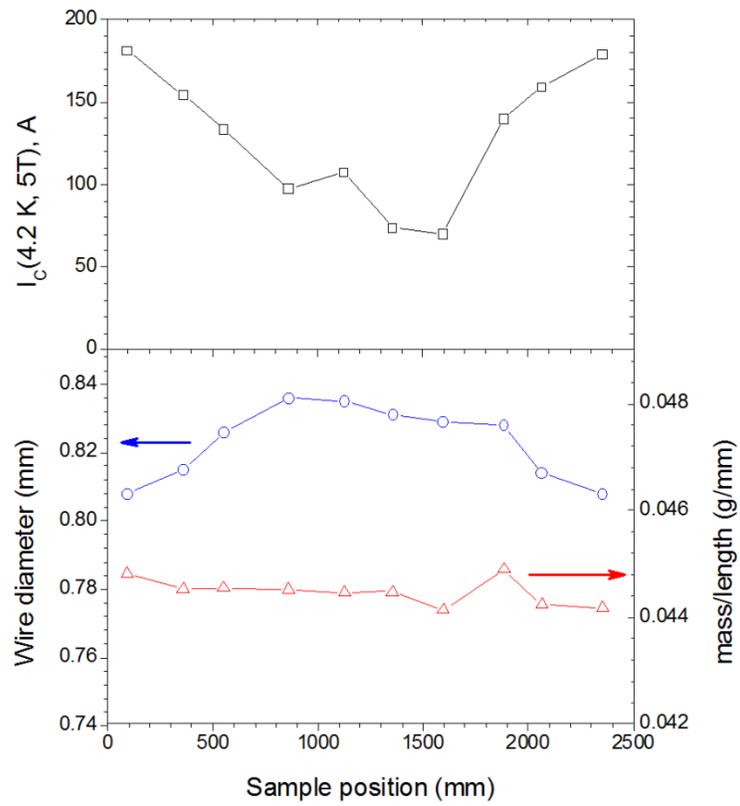

**Figure 3.** The variation of $I_C$ (4.2 K, 5 T), wire diameter and mass/length measured along the length of the 2.4 m open-ended wire. The wire diameter was calculated from the transverse cross section area measured by image analysis

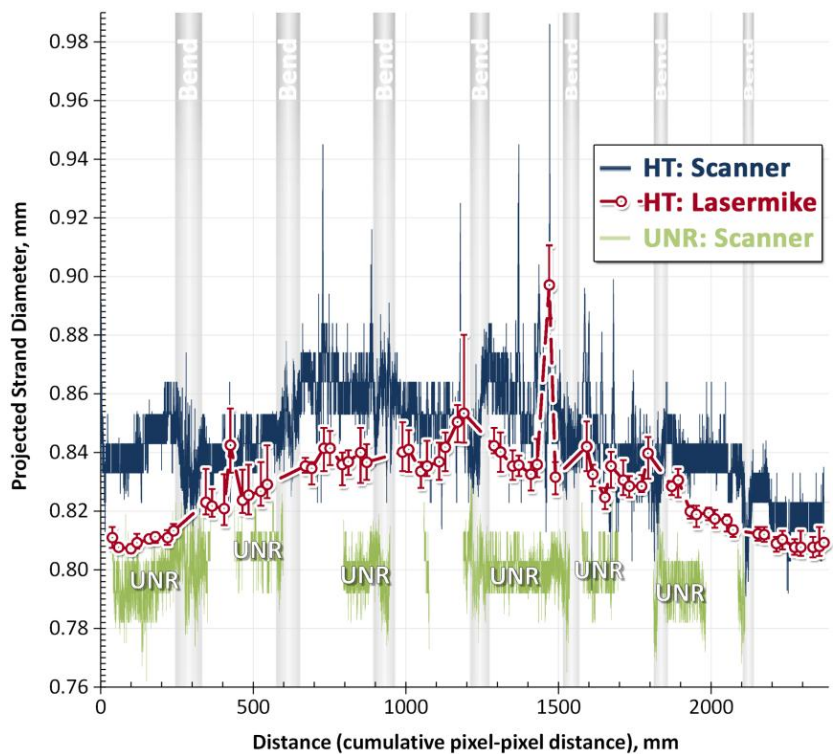

**Figure 4.** Wire diameter as a function of position along the entire length of the 2.4 m long open-ended wire before (UNR) and after full HT (HT). The diameter was measured with the flatbed scanner and the laser micrometer. The position of the bends where data could not always be obtained is indicated by the vertical grey bars.

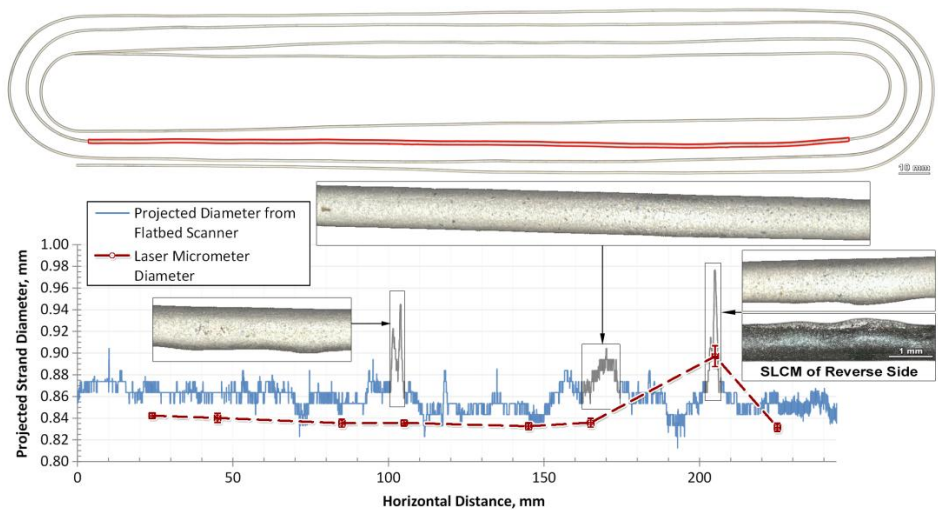

**Figure 5**. Wire diameters measured by the laser micrometer (dashed guide line) and the flatbed scanner (continuous line) for a single straight section indicated by the red-bordered length in the scanner image shown above of the open-ended 2.4 m strand. Also shown are images of specific sections of the wire with bulges and leaks taken with the flatbed scanner and the SLCM. The SLCM image on the lower right shows a double-bulge and a split in the strand where leakage occurred. This section of wire corresponds to the range from ~1250 to 1500 mm in Figure 4. It may appear that there is less diameter variation in Figure 4 than in Figure 5, but this is only because the diameter axis is expanded in Figure 4 to maximize detail of the diameter changes between the not-heat treated and the heat treated strands.

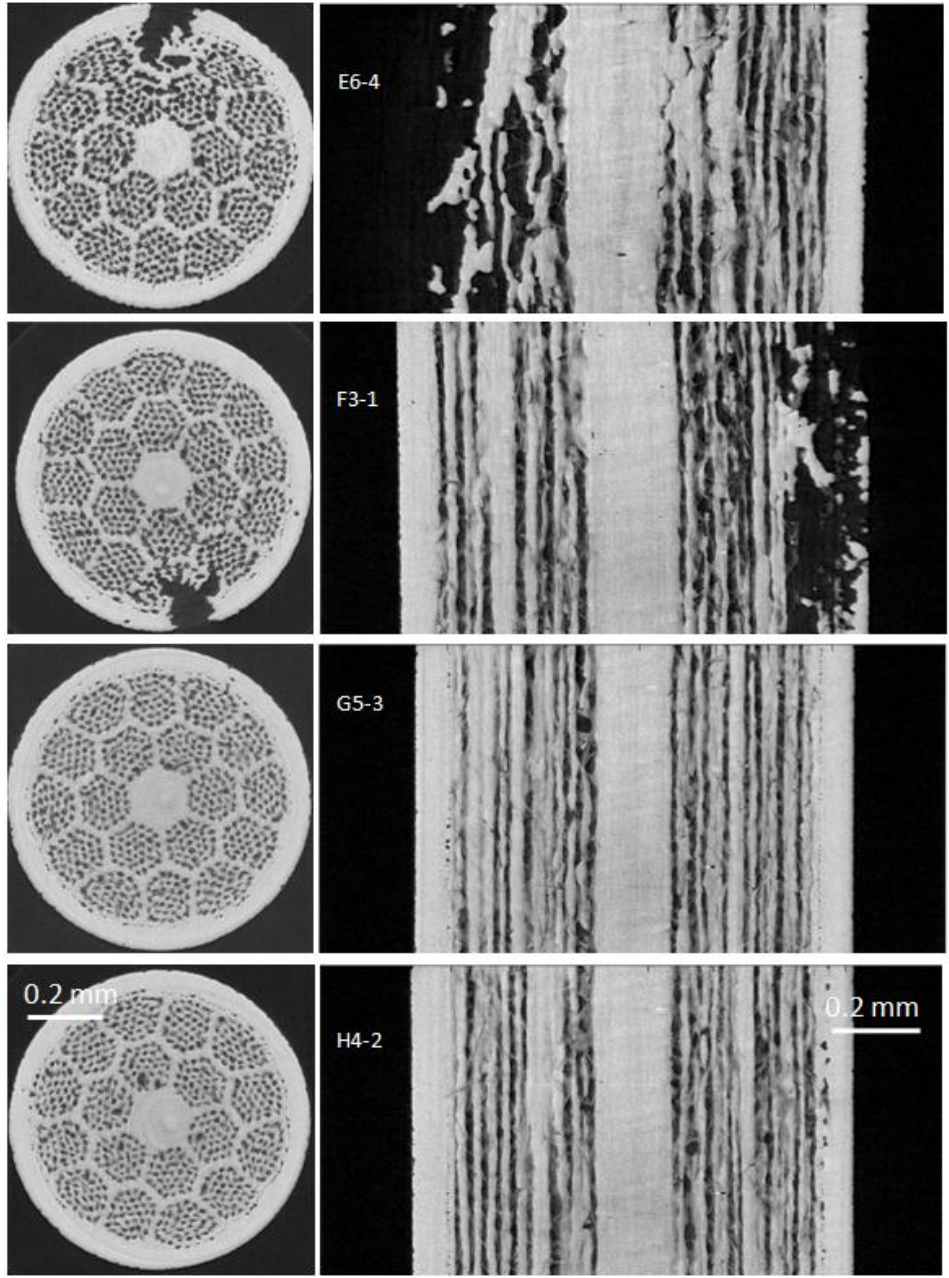

**Figure 6**. Transverse (left panel) and longitudinal (right panel) tomographic cross sections of sections of the 2.4 m open-ended wire cut at approximate distances from the end of 1480 mm (E6-4), 1780 mm (F3-1), 2000 mm (G5-3) and 2290 mm (H4-2).

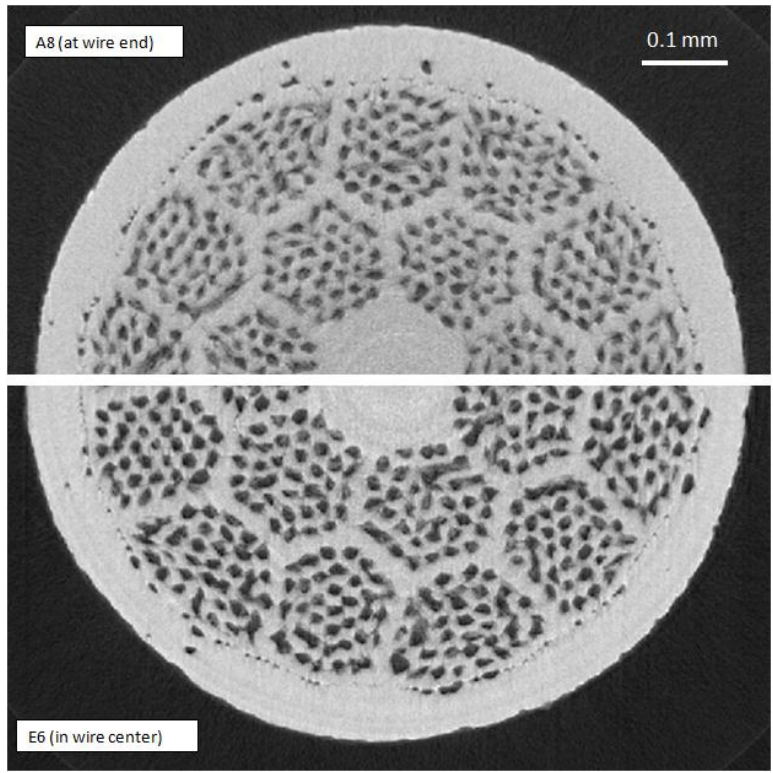

**Figure 7**. Two transverse tomographic half cross sections from one end (top) and near the midpoint (bottom) of the 2.4 m open-ended wire. The midpoint section clearly shows a larger diameter than the end of the wire. Voids in the filaments are gray to black. The greater void content of the filaments in the midpoint of the wire is indicated by their higher black density.

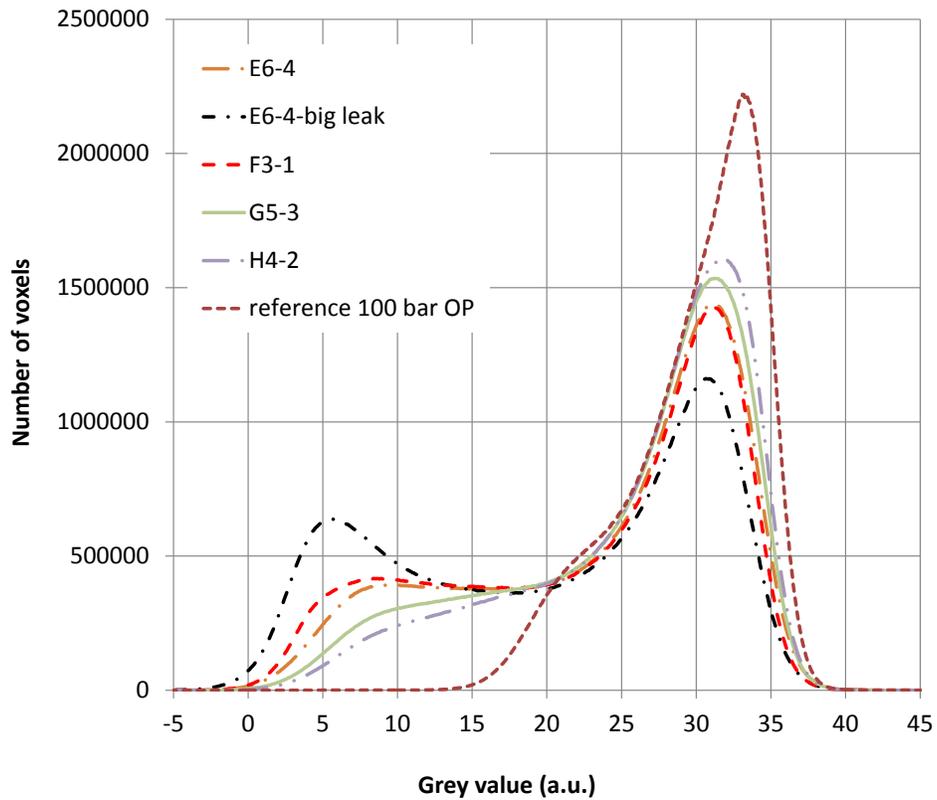

**Figure 8**: Histograms for section of wire cut from the 2.4 m open-end wire. A histogram from a wire where the porosity was strongly reduced by 100 bar overpressure processing is included as reference [35]. The strong variation in gray scale corresponding to the strongly varying mass density is clear. The large voids produced by the leak in Figure 7 produce a large peak around grey scale 5 while Ag produces grey scale values peaking around 31-33. The biggest effect is clearly on the filament grey scale density which extends over the large range of about 0 to 20. The tomography clearly indicates a very large range of 2212 density depending on the distance from the end of the sample.

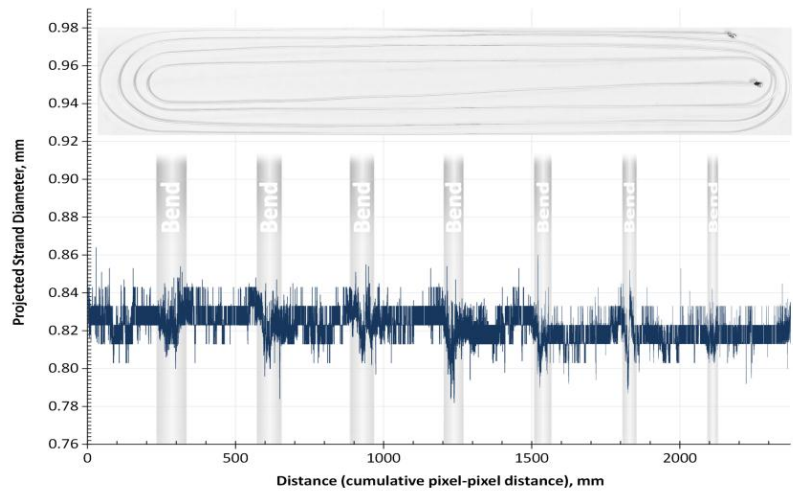

**Figure 9.** The top image shows the closed-end 2.4 m strand after heat treatment at 875°C below the melting point. The lower image shows that the wire diameter (measured with the flat-bed scanner) had expanded rather uniformly by 20-30 µm as compared to the 0.80 mm diameter as-drawn wire.

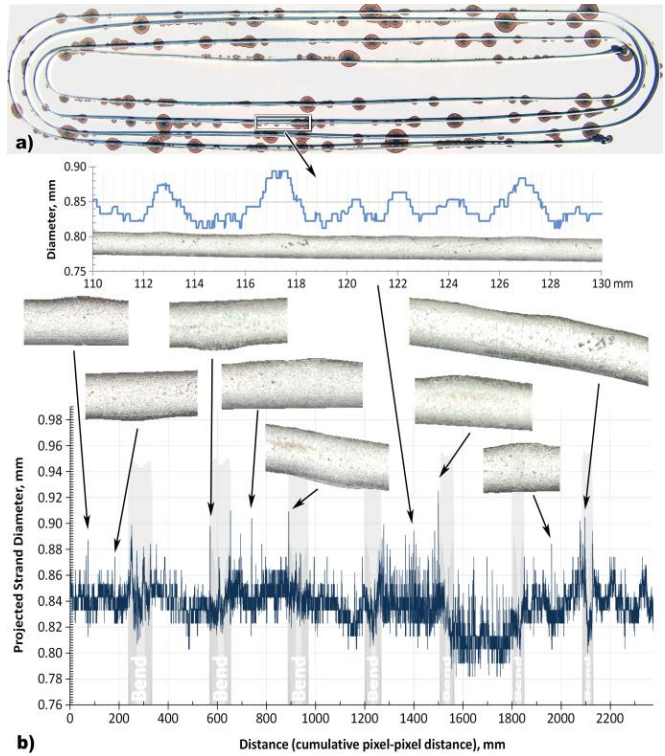

**Figure 10.** The top image shows a photograph of the closed-end 2.4 m strand after a 875°C HT (see Fig. 9) followed by a standard full melt heat treatment at 888°C. Melting produced significant leakage along the entire length of the wire and a highly variable diameter expansion. Images taken with the flatbed scanner of several sections of wire that show bulges are shown.